\documentclass[%
 reprint,
 amsmath,amssymb,
 aps,
 pra,
]{revtex4-2}

\usepackage[T1]{fontenc} 

\usepackage{graphicx} 
\usepackage{dcolumn} 
\usepackage{bm} 

\usepackage{amsmath}
\usepackage{amssymb}
\usepackage{amsfonts}
\usepackage{color}
\usepackage{ulem}

\newcommand*\diff{\mathop{}\!\mathrm{d}}

\begin{document}

\preprint{APS/123-QED}

\title{Vortices in the Many-Body Excited States of Interacting Bosons in Two Dimension} 

\author{Mateusz \'{S}lusarczyk}
 \email{slusarczyk.mat@gmail.com}
\author{Krzysztof Paw{\l}owski}
 \email{pawlowski@cft.edu.pl}
\affiliation{Center for Theoretical Physics, Polish Academy of Sciences, Al. Lotnik\'{o}w 32/46, 02-668 Warsaw, Poland}

\date{\today}

\begin{abstract}
Quantum vortices play an important role in the physics of two-dimensional
quantum many-body systems, though they  usually are understood in the
single-particle framework like the mean-field approach. Inspired by the study on
the relations between solitons and yrast states, we investigate here the
emergence of vortices from the eigenstates of a $N$-body Hamiltonian of
interacting particles trapped in a disk. We analyse states that appear by
consecutively measuring the positions of particles. These states have densities,
phases, and energies closely resembling mean-field vortices. We discuss
similarities, but also point out purely many-body features, such as vortex
smearing due to the quantum fluctuation of their center. The {\it ab initio}
analysis of the many-body system, and mean-field approaches are supported by the
analysis within the Bogoliubov approach. 
\end{abstract}

\maketitle
\begin{figure}[ht]
    \centering
    \includegraphics[width=8.6cm]{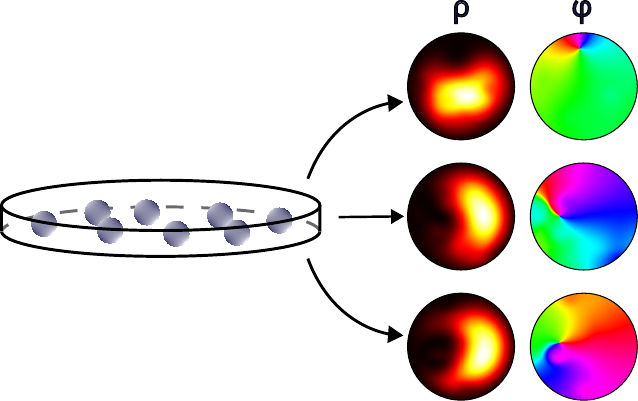}
    \caption{Graphical abstract: We study the many-body wave function of interacting bosons on a disk to detect vortices that become apparent after position measurement.}
    \label{fig:graphical-abstract}
\end{figure}

\section{Introduction}
Mean-field (MF) approaches are often the only way to tackle quantum many-body
systems and characterise their non-linear effects. At the same time, the
interpretation of the MF results in terms of the underlying exact model is
sometimes difficult and counter-intuitive. For instance, non-linear effects such
as solitons or vortices appear naturally in mean-field frameworks as solutions
of dynamical, time-dependent, single-particle, and non-linear mean-field
equations. On the other hand, the {\it ab initio} model is the linear, many-body
Schr\"odinger equation. Furthermore, it has been shown that the gray solitons,
constantly moving non-linear waves known from the MF approach, are related to
the non-moving many-body states, which are actually eigenstates of the many-body
Hamiltonian~\cite{Kulish1976, ishikawa1980}.

The relation between mean-field solitons and the eigenstates of the many-body
Lieb-Liniger model \cite{Lieb1963, LiebLiniger1963}, so-called yrast states
\cite{Mottelson1999} was a subject of many papers \cite{Sato2012, syrwid2015,
syrwid2016, golletz2020, Brand2019, Kaminishi2020Apr, KanamotoCarr2008,
Kaminishi2011, Fialko2012}. In between, it has been found that dark soliton,
robust dip in a gas density, emerge via breaking symmetry of the system
initiated in the yrast state. The symmetry may be broken by simply ``measuring''
part of the system \cite{syrwid2015, syrwid2016}, which in cold gases will
happen spontaneously due to particle losses \cite{golletz2020}. On the other
hand, if the gas is prepared in a product state, with all atoms occupying the
mean-field soliton, one finds within Bogoliubov theory quantum correction to
this state -- the black MF soliton has quantum fluctuations of its position
\cite{Sacha2002} due to anomalous negative mode. The effect of these
fluctuations is to gray out the soliton, even though it appears black in the MF
description.

The aim of this paper is to put the research further into a study of emergence
of 2D  vortices out of the many-body yrast states. Quantum vortices, understood
via MF approach, play an important role in many contexts, demonstrated in a
number of experiments, on rotated superfluid \cite{Madison2000}, on vortex
lattice \cite{Coddington2004, AboShaeer2002,Engels2002}, quantum turbulent gas
\cite{HennBagnato2009} or the Berezinsky-Kosterlitz transition
\cite{Hadzibabic2006Jun} (see \cite{Tempere17} for more information on vortices
in ultracold quantum gases). Ongoing research on vortices in 3D turbulence is
making use of AI techniques \cite{keepfer2023, Metz_2021, Kim_2023} for
analyzing numerical \cite{Metz_2021},  and experimental results \cite{Kim_2023},
assuming mean-field description. While mean-field methods are commonly employed
in these studies, it is essential to acknowledge their limitations. The
observation of unexpected quantum droplets \cite{Kadau_2016}, which arise due to
the quantum corrections \cite{Petrov2015}, highlights the need to further study
model systems at the quantum many-body level and compare them with the MF
approach.

In the present work, we focus on a gas of atoms with short-range interactions
inside a 2D circular plaquette (see Fig.~\ref{fig:graphical-abstract}). Unlike
in 1D, there are no analytical solutions for vortices -- neither in the
many-body  nor in the MF approach. In fact, even numerically, it is not
straightforward to find such dynamical solutions to the MF, that would posses a
single vortex with core outside the center of the trap. Our  starting points are
the many-body yrast states  -- states minimizing energy at fixed momentum, which
we find using the exact diagonalization method  with importance truncation
\cite{Roth2009}. We show that during consecutive ``measurements'' of positions
of particles initiated in an yrast state, one reveals vortices. They may appear
off the plaquette center, and at random positions, but are robust dynamically,
as confirmed by evolving it with the mean-field equations. In a sense, our
complicated procedure serves approximate solution to the MF approach. Special
attention is paid to states with angular momentum $j\hbar$, with integer $j$ not
larger than the total number of atoms $ N$. We show that if $j=N$, the position
of the vortex revealed from the quantum states fluctuates. As a result, the
average density profile differs between the MF and many-body approach, even in
this case. The differences are attributed to the quantum fluctuations, as
confirmed by the Bogolubov analysis, similarly to the case of "graying" of the
1D  black soliton \cite{Sacha2002}. Most of the attention is paid to the case
$j<N$, in which a vortex emerges out of the center.

In Sec. \ref{sec:model}  we present our model. Sec. \ref{sec:GS} is devoted to
the ground state, for which MF gives very good results, for as few as $6$ atoms.
In Sec. \ref{sec:YS} we show that vortices are not apparent in the densities,
but they emerge in a stochastic process mimicking measurements on many-body
excited states. We conclude in Sec \ref{sec:conclusions}.

\section{Model\label{sec:model}}
We illustrate our findings for 2D vortices on an example of system of bosons
interacting via short range potential and trapped inside a quasi-2 dimensional
disk, i.e. assuming infinite potential outside a disk of radius $R$ and
neglecting the $Z$ direction. Our Hamiltonian of interest in the second
quantized form reads 
\begin{align}
    \label{eq:H_field}
    \hat{H} & = \int \hat{\psi}^\dag  ({\bm r}) \left( -\frac{\hbar^2}{2M} \nabla^2\right) \hat{\psi}  ({\bm r}) \diff{\bm r} \\
    \nonumber
    & + \frac{1}{2}\iint \hat{\psi}^\dag ({\bm r}) \hat{\psi}^\dag ({\bm r^{\prime}}) V({\bm r}-{\bm r^{\prime}}) \hat{\psi}({\bm r^{\prime}}) \hat{\psi} ({\bm r})\, \diff{\bm r}\diff{\bm r^{\prime}},
\end{align}
where $M$ is the atomic mass, $\bm{r} = [x,\,y]$ is a two-dimensional position
and  $\hat{\psi}(\bm{r})$ is the field operator obeying the bosonic commutation
relations $[\hat{\psi}(\bm{r}), \,\hat{\psi}^{\dagger}(\bm{r}')] =
\delta(\bm{r}-\bm{r}')$. Here we model the interaction potential with a
repulsive Gaussian $V({\bm r})= \frac{g}{\pi \sigma^2} \,
e^{-\bm{r}^2/\sigma^2}$, where $g$ is the interaction coupling constant. In what
follows we assume that the interaction range $\sigma$ is significantly smaller
that the average distance between atoms. The key advantage of using a Gaussian
potential, instead of the usual Dirac delta, lies in its convergence -- unlike
in  the Dirac delta potential case, the many-body problem remains well defined
without requiring regularization \cite{Albeverio1988} or a suitable adjustment
of the interaction strengths \cite{Rontani_2017}. Moreover, for the Gaussian
potential, one can evaluate the 2D scattering length analytically to some extent
\cite{Jeszenszki2018}.

On the other hand, even for the smooth Gaussian interaction potential, the
numerics for $N$ particles are involved (see discussion in
\cite{Jeszenszki2019}). We have achieved satisfactory convergence for $N=6$
atoms using the importance truncation method \cite{Roth2009}. Below we discuss
without details our numerical approach, whereas the convergence tests are
discussed in the \ref{ap:convergence}.

Firstly, we decompose the quantum field operator in a single-particle basis as follows:
\begin{equation}
    \label{eq:psi_operator}
    \hat{\psi}(r,\phi) = \sum_j \psi_j(r,\phi)\, \hat{a}_{j},
\end{equation}
where index $j$ runs over all single-particle states. For our trap geometry
those are given as products of  radial part and a phase factor
\begin{equation}
    \label{eq:single_particle_state}
    \psi_j(r,\phi) = f_{j}(r) e^{im_j\phi},
\end{equation}
with radial part being proportional to the Bessel function of the first kind
\begin{equation} \label{eq:radial_sp_wave_fun}
    f_{j}(r) = A_j J_{m_j} \left( \alpha_j \frac{r}{R} \right).
\end{equation}

\begin{figure}[ht]
    \centering
    \includegraphics{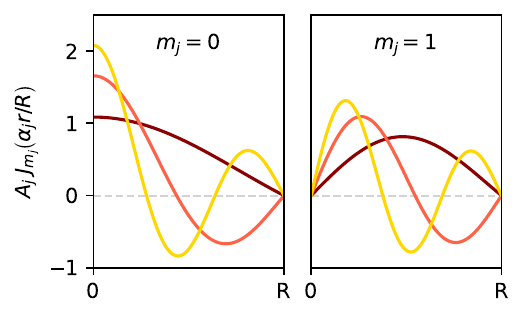}
    \caption{
        Examples of radial parts of the single particle states used in numerics,
        Eq.~\eqref{eq:radial_sp_wave_fun} for $m_j = 0$ (left) and $m_j=1$
        (right). All radial parts for $m_j = 0$ are the $0$th Bessel functions,
        but rescaled by $\alpha_j$ (with $\alpha_j$ being zeros of the $0$th
        bessel function). Similarly, all radial parts for  $m_j = 1$ equal to
        rescaled  1st Bessel function.
    }
    \label{fig:bessels}
\end{figure}

For convenience, we label the single-particle eigenstates with a multi-index $j$
that encodes both the magnetic quantum number $m$ and the quantum number
labeling the radial eigenstates. The integer $m_j$ represents the angular
momentum of a given state in units of $\hbar$. For a given $m_j$, all radial
excitations are described by the first-kind $m_j$-th Bessel functions, but
rescaled by the factor $\alpha_j$. The scaling factors $\alpha_j$, which
distinguish different radial excitations for a given $m$, are consecutive zeros
of the $m_j$-th Bessel function. This construction guarantees that the boundary
condition $\psi_j(R, \phi)=0$ is always satisfied. Finally, $A_j$ is the
normalization constant. Examples of the radial part of the wavefunctions are
given in Fig.~\ref{fig:bessels}. Energy of such a single particle state follows
the formula
\begin{equation}
    E_j = \frac{\hbar^2\alpha_j^2}{2M R^2}.
\end{equation}

Inserting (\ref{eq:psi_operator}) into (\ref{eq:H_field}) we obtain the
Hamiltonian in terms of the creation and annihilation operators

\begin{equation}
    \label{eq:H_a}
        \hat{H} = \sum_{j}E_j\hat{a}_j^\dag \hat{a}_{j} + \frac{1}{2}\sum_{ijkl} V_{ijkl} \delta^{m_i+m_j}_{m_k+m_l} \hat{a}_{i}^\dag \hat{a}_{j}^\dag\hat{a}_{l} \hat{a}_{k},
\end{equation}

where $V_{ijkl}$ is a double integral
\begin{eqnarray}
    \label{eq:V_ijkl}
    V_{ijkl} = && \int_0^R \int_0^R r r' f_i(r) f_j(r') f_k(r) f_l(r') \nonumber \\
    && \times \frac{4\pi g}{\sigma^2}e^{-\frac{r^2 + r'^2}{\sigma^2}} I_{|m_i-m_k|}\left( \frac{2rr'}{\sigma^2} \right) \diff{r} \diff{r'},
\end{eqnarray}
where $I_n(x)$ is the n-th modified Bessel function of the first kind. Note that
the Hamiltonian is invariant under rotation, which implies that angular momentum
is conserved. As a result, the Kronecker delta appears in Eq.~\eqref{eq:H_a}.
This invariance also implies that all eigenstates of the Hamiltonian can be
chosen to be also eigenstates of the angular momentum operator $\hat{L}_z$. 

Our objects of interest are the excited states of the above
Hamiltonian~\eqref{eq:H_a}, that minimize the energy at a fixed angular
momentum, so called yrast states \cite{Mottelson1999}. In our paper we will
compare them with the states arising in the simplistic description of the
system,  the Gross-Pitaevskii equation (GPE), and the Bogoliubov approximations.

{\it Mean-field approximation} 
In the MF approach, it is usually assumed that every particle resides in the
same state $\psi_\mathrm{GP}(\bm r, t)$ and its dynamics are governed by the GPE
\begin{equation}
    i\hbar\frac{\partial}{\partial t}\psi_\mathrm{GP} = 
    \left( -\frac{\hbar^2}{2M} \nabla^2 + (N-1) V_{\rm MF} ({\bm r}, t) \right) \psi_\mathrm{GP},
\end{equation}
where the mean-field potential $V_{\rm MF}$ is given by the convolution between
the Gaussian interaction potential $V$ and the mean-field density
\begin{equation}
V_{\rm MF} ({\bm r}, t) = \int  V({\bm r}-{\bm r}^{\prime})\left|\psi_{\rm GP}({\bm r^{\prime}}, t)\right|^2\diff{\bm r^{\prime}}.
\end{equation}
We also compare the many-body results with the solutions of the stationary GPE:
\begin{equation}
    \left( -\frac{\hbar^2}{2M} \nabla^2 + (N-1)V_\mathrm{MF}(\bm r) \right) \psi_\mathrm{GP} = \mu\, \psi_\mathrm{GP}.
    \label{eq:GPE}
\end{equation}

Imposing the constrain $\psi_\mathrm{GP}(\bm r) = f(r)e^{i\phi}$ yields a
solution with single vortex at the center. In contrast, the ground state can be
found assumming $\psi_\mathrm{GP}(\bm r) = f(r)$, which depends only on the
distance from the center.

The GPE is the simplest version of the mean-field theory for our system -- via
non-linear term a single particle ``feel'' all other particles due to effective
potential $V_{\rm MF}$ created by all other particles. We will not invoke in
this paper more elaborated mean-field versions, like the ones based on the
density-functional theories or generalizations based on the LHY corrections.

Instead we will invoke the Bogoliubov approximation -- many-body approach, but
based on the assumption that almost all atoms occupy the MF orbital $\psi_{\rm
GP}$, with only a small fraction of atoms in other single-particle states
\cite{Castin2001May, Castin2002Jul}. Technically, one has to find the Bogoliubov
modes $u(\bm{r})$ and $v(\bm{r})$ that obey the Bogoliubov-de Gennese equations:

\begin{equation}
        \begin{pmatrix}
                \hat{H}_\mathrm{GP} + \psi_\mathrm{GP}\hat{U}^* & \psi_\mathrm{GP}\hat{U} \\
                -\psi_\mathrm{GP}^{*}\hat{U}^* & -\hat{H}_\mathrm{GP} - \psi_\mathrm{GP}^*\hat{U}
        \end{pmatrix}
        \begin{pmatrix}
                u_n \\
                v_n
        \end{pmatrix} =
        \epsilon_n
        \begin{pmatrix}
                u_n \\
                v_n
        \end{pmatrix},
        \label{eq:bdg}
\end{equation}
where 
\begin{equation}
    \hat{H}_\mathrm{GP} =  -\frac{\hbar^2}{2M} \nabla^2 + (N-1)V_\mathrm{MF}(\bm r) - \mu.
\end{equation}
We introduce linear operator $\hat{U}$, which depends on $\psi_\mathrm{GP}$, and
whose action on $u(\bm r)$ is given by
\begin{equation}
    \hat{U}u(\bm r) = (N-1)\int V(\bm r - \bm r') \psi_\mathrm{GP}(\bm r') u(\bm r') \diff{\bm r'}.
\end{equation}

Solutions can be classified into three families: the ``$+$ family'', the ``$-$
family'', and the ``0 family''. These names correspond to the normalization of
the expression $\langle u_n|u_n\rangle - \langle v_n|v_n\rangle$, which equals
$+1$, $-1$ and $0$, respectively \cite{Castin2001May}. Having $u_n (\bm{r})$ and
$v_n(\bm{r})$ one is in position to compute quantities to characterize the
system, as the ones listed below.

{\it Quantities} 
In our comparisons between the many-body and GPE approaches, we compute energies
and density profiles and then track the differences with the help of the
correlation functions. 

The energy of the many-body yrast states comes simply from exact diagonalization
of the hamiltonian \eqref{eq:H_a}. It is compared, wherever possible, with the
mean-field energy:
\begin{align}
    E_{\rm GP}/N = &  -\frac{\hbar^2}{2M} \int\psi^*_{\rm GP} \nabla^2 \psi_{\rm GP} \diff{\bm r}  \\
      + \frac{N-1}{2} & \iint \left|\psi_{\rm GP}({\bm r})\right|^2 V({\bm r^{\prime}}-{\bm r})\left|\psi_{\rm GP}({\bm r^{\prime}})\right|^2\diff{\bm r^{\prime}} \diff{\bm r},\nonumber
\end{align}
and the ground state energy approximated as in the Castin-Dum approach
\cite{Castin1998} used here to the non-local interaction potential \footnote{The
formula for the ground state energy in Bogoliubov \eqref{eq:EBdG} was
\cite{Castin1998} in the context of the short range interaction. We apply it in
the same form to our problem.}:
\begin{equation}
    \label{eq:EBdG}
    E_\mathrm{BdG} = E_\mathrm{GP} - \sum_{n\in\mathrm{``+family''}} \epsilon_n \langle v_n | v_n \rangle.
\end{equation}

Another figure of merit is the gas density. In MF the density is simply
$N|\psi_{GP}|^2$.  For many-body system, one has to compute  the single particle
density matrix (SPDM)
\begin{equation}
    \label{eq:rho}
    \rho(\bm r, \bm r ') := \langle \Psi | \hat{\psi}^\dag(\bm r ') \hat{\psi}(\bm r) | \Psi \rangle.
\end{equation}
Its diagonal, $\rho (\bm r) = \rho(\bm r, \bm r)$, yields information about the
average density of particles in the many-body state $|\Psi\rangle$.

The SPDM can be used to test the assumptions underlying MF and BdG approaches,
that have the validity range restricted to the weakly interacting limits.
Diagonalization of SPDM yields the expansion
\begin{equation}
    \label{eq:rho_decomposition}
    \rho(\bm r, \bm r ') = N\sum_i \lambda_i \psi_i^*(\bm r')\psi_i(\bm r),
\end{equation}
where sum of the coefficients $\lambda_i$ is equal to $1$. Regime of weak
interactions corresponds to situations where majority of the system lies in a
single state, say $\psi_0$, with its eigenvalue $\lambda_0$ being close to $1$,
that is $\lambda_0 \gg \lambda_i$ for $i \neq 0$.

The value of $\lambda_0$ can be compared to the MF fraction $(N- \langle \delta
\hat{N} \rangle)/N$ of particles in condensate, where $\langle \delta \hat{N}
\rangle$ is the average number of particles out of condensate and can be
calculated from BdG solution using formula \cite{Castin2001May, Castin2002Jul}
\begin{equation}
    \langle \delta \hat{N} \rangle \simeq \sum_{n\in\mathrm{``+family''}}\langle v_n | v_n \rangle.
\end{equation}

\section{Ground state \label{sec:GS}}
\begin{figure}
    \centering
    \includegraphics{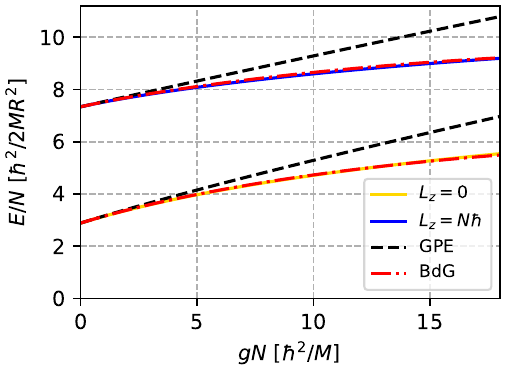}
    \caption{
        Energy per atom for states with $N=6$  particles as a function of
        coupling constant $g$ obtained from the exact diagonalization (solid
        lines), GPE approach (black dashed lines) and Bogoliubov approach (red
        dot-dashed lines). The lower curves correspond to the ground state, i.e.
        $0$ total angular momentum, whereas the upper ones to the states
        minimizing energy with total angular momentum equal to the total number
        of particles. All results are presented for $\sigma=0.1\ R$.
    }
    \label{fig:energies}
\end{figure}

To ensure that MF and MB (many-body) approaches are mutually consistent, we
begin by investigating the ground state. In Fig.~\ref{fig:energies} we show
energy of ground state obtained from exact diagonalization (yellow line)
calculated for different coupling constants $g$. In the regime of weak
interactions, this result of the exact diagonalization  overlaps with results
from the mean-field GPE calculations (black line) up to interactions around
$gN\approx 5$. The MF energy as proportional to $g$, is just a straight line
that gives a correct slope to the exact energy at $g=0$ but it does not capture
bending off the energy curve. The latter behaviour is well reproduced by the
energy computed within the BdG approach (dot-dashed line) in the whole
investigated range of interaction strengths $g$. 

\begin{figure}[ht]
    \includegraphics{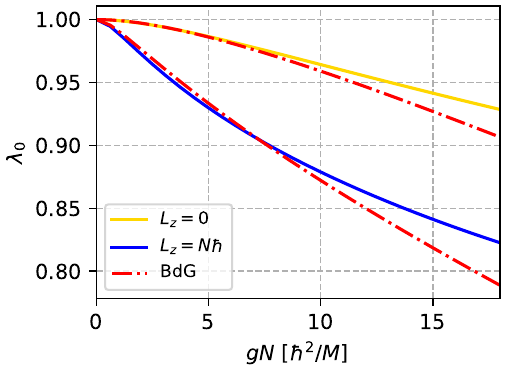}
    \caption{
        Condensed fraction of a system described by coefficient $\lambda_0$ from
        equation \eqref{eq:rho_decomposition} as a function of the interaction
        strength. Calculations were performed for $N=6$ particles in the ground
        state (yellow line) and the yrast state with angular momentum $N\hbar$ (blue
        line). The red dot-dashed lines illustrate the results obtained from the
        Bogoliubov approach.
    }
    \label{fig:depletion}
\end{figure}

To illustrate further the consistency between different approaches we check
whether indeed until $gN \approx 5$, the assumption underlying MF and BdG is
fulfilled, i.e. if the system is dominated by a single mode. Data for such
analysis is illustrated in  Fig.~\ref{fig:depletion} showing the relative
occupation of the dominant orbital, i.e.  $\lambda_0$ as defined in Eq.
\eqref{eq:rho_decomposition}. As expected, up to $gN \approx 5$, practically all
atoms reside in a single orbital in the ground state.

\begin{figure}[ht]
    \includegraphics{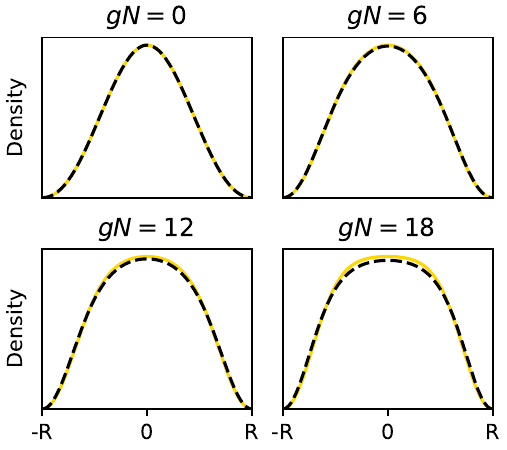}
    \caption{
        Ground state density of particles obtained from MF calculations (black
        dashed) and SPDM from exact diagonalization (yellow solid). Calculations
        were performed for $N=6$ particles, the interaction strength $gN$ is
        given in units $\hbar^2/M$. Up to the largest considered value $gN=18$,
        both curves can be seen to overlap. The range of the Gaussian
        interaction potential equals to $\sigma = 0.1\ R$.
    }
    \label{fig:gpe_mb_rho_m0}
\end{figure}

To complete the analysis of the ground state we also compare  density profiles,
see Fig.~\ref{fig:gpe_mb_rho_m0}. It shows the cross-section of MF solution
$N|\psi_\mathrm{GP}|^2$ and the diagonal of SPDM, $\rho(\bm r)$, for some
selected values of the interaction strength $g N$. The density profiles are very
similar, even for substantially large interaction strengths. 
To conclude this part -- in the weakly interacting regime, i.e. as long as the
depletion of the condensate remains small ($\lambda_0 \approx 1$), the energy
and the density profiles are well predicted by the MF approach. In this case,
comparing densities is straightforward: the single-particle density obtained
from the density matrices of the many-body ground states agrees with the MF
densities $N|\psi_{\rm GP}|^2$. This indicates a lack of substantial quantum
correlations between atoms. In the following, we discuss the yrast states for
which the situation is much more subtle.

\section{Yrast states} \label{sec:YS}
We now turn to the analysis of the eigenstates minimizing  energy at a fixed
angular momentum, so called yrast states. We will denote the quanta of angular
momenta with integer $l$, which will correspond to the value of the angular
momentum equal to $l\hbar$. The case of $l=0$ corresponds to the ground state of
the system described in the preceding section. We expect that yrast states with
$l\neq0$ contain quantum vortices.
\begin{figure}[ht]
    \includegraphics{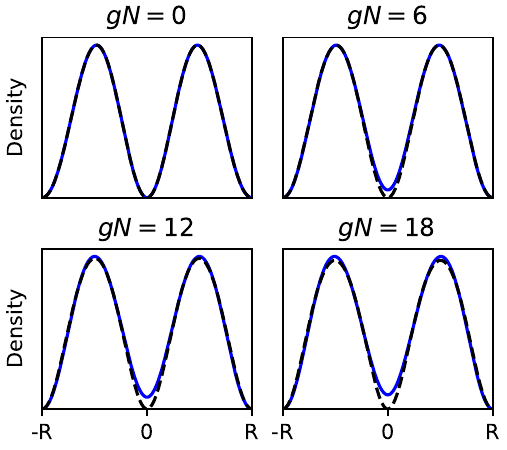}
    \caption{
        Yrast state density of particles obtained from MF calculations (black
        dashed) and SPDM from exact diagonalization (blue solid). Calculations were
        performed for $N=6$ particles and angular momentum $L_z=N\hbar$, the
        interaction strength $gN$ is given in units $\hbar^2/M$.
    }
    \label{fig:gpe_mb_rho_m6}
\end{figure}
We begin our discussion with the special case $l=N$, where the angular momentum
is evenly distributed among all the particles, each carrying $\hbar$. Then, each
atom shall rotate with angular momentum $\hbar$ around the disc center, and all
atoms altogether will form a single vortex. Similarly to the $l=0$ case, we
compare energies (Fig.~\ref{fig:energies}), the quantum depletion
(Fig.~\ref{fig:depletion}), and the densities (Fig.~\ref{fig:gpe_mb_rho_m6})
calculated using the {\it ab initio} and MF approaches. Unlike in the ground
state case, as the interaction strength increases, differences in results of
these two approaches become apparent. The quantum depletion grows much quicker
with interaction, as compared with the ground state case. This implies
discrepancies in the gas densities. Although the density cross-sections look
very similar in both approaches, the MF densities deviates from the SPDM in the
center of the system -- the density at the center always vanishes in the MF
approximation, whereas it is more and more filled with matter in the fully
quantum analysis, at least if it comes to the diagonal of a single particle
density matrix. We will postpone for a while discussion of our interpretation of
the quantum case. The Reader may know similar effect, from the physics of
vortices in fermionic superfluid \cite{Caroli1964May} or from the discussion of
graying bosonic solitons due to anomalous Bogoliubov mode
\cite{Sacha2002,Dziarmaga2003Mar}.

\begin{figure}[ht]
    \includegraphics{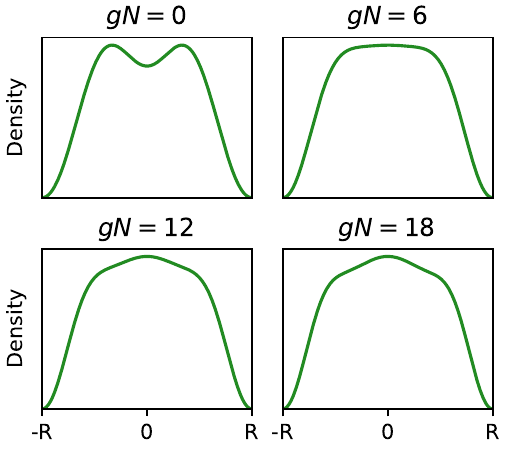}
    \caption{
        The single-particle density (SPDM) $\rho(\bm{r})$ as a function of a
        position of a system in an yrast states with angular momentum $L_z =
        N\hbar/2$, obtained via exact diagonalization. There is no clear indication
        of where the vortex might be located. Parameters: $N=6$ particles,  angular
        momentum $L_z = 3\hbar$. The interaction strength $gN$ is expressed in units
        of $\hbar^2/M$.
    }
    \label{fig:gpe_mb_rho_mhalf}
\end{figure}

More puzzling is the case $0<l<N$.  In this case, different scenarios are
possible to imagine --  off-centered vortex or vortices, with matter rotating
around a vortex core, which itself rotates around the center of a trap.  It is
no straightforward to find numerically such GPE solutions - we haven't done it.
Instead, we study this case starting from the many-body approach -- finding
yrast state using exact diagonalization method. Our goal here is to show  that
MF vortices are encoded in the yrast states, and can be revealed via
measurements. Firstly, we show the SPDM for $l=N/2$ in
Fig.~\ref{fig:gpe_mb_rho_mhalf}. This SPDM  does not give a clear view of the
system. In particular it shows no vortices. The analysis for the yrast states at
a fixed angular momentum $0<l\hbar<N\hbar$ looks as follows: (i) one cannot find
easily these states using MF (ii) the density obtained from the single-particle
reduced density matrix, Eq. \eqref{eq:rho}, change substantially with
interaction and does not show any vortex/vortices. Therefore we use a more
subtle approach using high-order correlation function, as described below.

To this end, we use a procedure similar to the one introduced to explain an
interference pattern arising in the system in a Fock state
\cite{Javanainen1996}, also used to find solitons within 1D yrast states
\cite{Dziarmaga2003Mar, syrwid2015}. The procedure relies on drawing
(``measuring'') positions one-by-one from MB state $|\Psi\rangle$ (see Appendix
\ref{ap:measurement} for details). One ends up with $N-1$ points $\bar{\bm
r}_1,\dots,\bar{\bm r}_{N-1}$, which then enter as parameters to the conditional
wave function
\begin{equation}
    \label{eq:conditional_wave_function}
    \psi_\textrm{cond}(\bm r) \propto \langle \bar{\bm{r}}_1, \ldots , \bar{\bm{r}}_{N-1}, \bm r | \Psi \rangle,
\end{equation}
that is dependent only on one position $\bm r$, but it is stochastic. Note that
$|\psi_\textrm{cond}(\bm r)|^2$ is proportional to the $N$-th order correlation
function 
\begin{equation}
    G_N =\langle 
    \hat{\psi}^\dagger (\bar{\bm{r}}_1) \ldots \hat{\psi}^\dagger (\bar{\bm{r}}_{N-1})\hat{\psi}^\dagger (\bm{r})\hat{\psi} (\bm{r})\hat{\psi} (\bar{\bm{r}}_{N-1})\ldots\hat{\psi}(\bar{\bm{r}}_{1})\rangle.
\end{equation}
We search for vortices within the conditional wavefunctions
$\psi_{\mathrm{cond}}$, or in other words -- within the high-order correlation
function.

\begin{figure}[ht]
    \centering
    \includegraphics{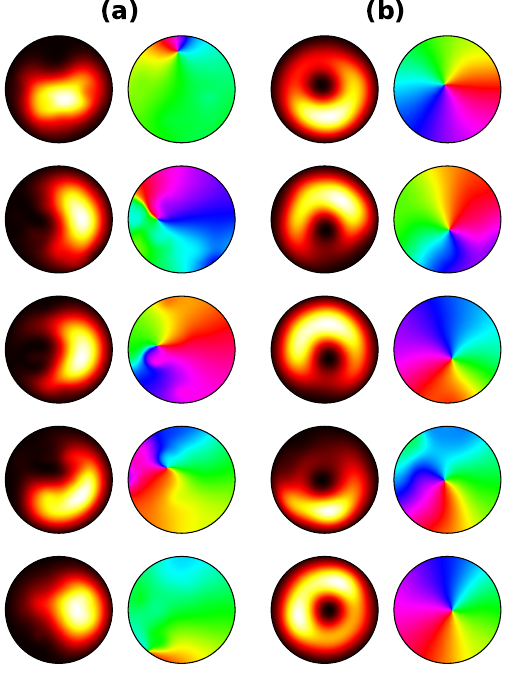}
    \caption{
        Density and phase of exemplary conditional wave functions
        \eqref{eq:conditional_wave_function} for $N=6$ particles, interaction
        strength $gN = 18\hbar^2/M$ and the total angular momentum $N\hbar/2 =
        3\hbar$ (a) and $N\hbar = 6\hbar$ (b). In both panels there are two columns
        -- the left one for the density and the right one for the phase of
        $\psi_{\rm cond}$ In both cases, a vortex, if appears, emerges off-centered,
        while in (a) the variation of its position is much greater.
    }
    \label{fig:draws_gN6}
\end{figure}

Few conditional wave functions for $N=6$ particles, $gN = 18\hbar^2/M$ and
$l=N/2$ are presented in Fig.~\ref{fig:draws_gN6} (a). In most cases one
observes a single vortex although its position differs between realisations. To
verify whether the vortex-like states emerging in the measuring procedure are
related to the MF vortices, we use $\psi_{\rm cond} (\bm{r})$ as initial states
for GPE equation and see their dynamics -- as in the movies for two interaction
strengths shown in the Supplemental Material  \footnote{See  Supplemental
Material.}. We found that the vortices are dynamically stable - they rotate
around the disc center almost not distorted for weaker interaction $gN =
6\hbar^2/M$. For stronger interaction $gN = 18\hbar^2/M$ the system has
non-trivial evolution -- effect of the beyond mean-field corrections. On other
the hand, the density averaged over many measurements, that correspond to the
SPDM, will show no vortices -- they will be hidden due to the large dispersion
in the random position of the vortex core. In the case $l=N$
(Fig.~\ref{fig:draws_gN6} (b))  the dispersion in the vortex position is much
smaller, consistent with a clear vortex in the disc center shown in the SPDM in
Fig.~\ref{fig:gpe_mb_rho_m6}, just slightly greyed-out. 

\begin{figure}[ht]
    \centering
    \includegraphics[width=6.4cm]{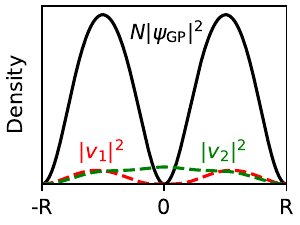}
    \caption{
        The cross section of the MF density $N|\psi_\mathrm{GP}|^2$ (black solid) is
        compared with the densities $|v_k|^2$ of two BdG modes that contribute the
        most to depletion, one with $L_z = 0$ (green dashed) and one with $L_z =
        2\hbar$ (red dashed). Computations were performed for $N=6$ and $gN =
        18\hbar^2/M$.
    }
    \label{fig:BdG_mode}
\end{figure}

The dispersion of the center of mass around the center of the disk for $l=N$ can
also be analyzed using the BdG approach. Numerically, we find the vortex with
$l=N$ using the GPE equation and then study its excitations using the BdG
equation. Our analysis reveals two modes that contribute most significantly to
the quantum depletion, each with roughly equal impact. The first of them
corresponds to depletion to $L_z = 0$ state. Located in the center, it is
responsible for filling the vortex core visible in the SPDM in
Fig.~\ref{fig:gpe_mb_rho_m6}. Its presence is also compatible with the random
vortex position in the conditional wave function $\psi_\mathrm{cond}$; the
distribution of these positions is likely related to the width of this mode. The
second mode, with $L_z = 2\hbar$, ensures the conservation of angular momentum.
The densities $|v|^2$ of these modes are shown in Fig.~\ref{fig:BdG_mode}, along
with the vortex profile.


\section{Conclusions} \label{sec:conclusions}
We studied a 2D gas in these excited states that minimize energy at a given
angular momentum, known as the yrast states. In a 1D version of this problem,
such states are related to mean-field solitons. We have shown that in the 2D
version, the states are related to vortices. The vortices  emerge spontaneously
during measurements, appearing in the high-order correlation functions of the
yrast states. When the total angular momentum is less that the number of
particles in the system, the system exhibits a single off-center vortex rotating
around the center. We demonstrate that the MF-like vortex arising in this manner
is always subject to center-of-mass fluctuations, as confirmed by Bogoliubov
analysis. The data that support the findings of this article are openly
available~\cite{slusarczyk_2025_15341582}.

\section{Acknowledgements}
Center for Theoretical Physics of the Polish Academy of Sciences is a member of
the National Laboratory of Atomic, Molecular and Optical Physics (KL FAMO).

K.P. acknowledge support from the (Polish) National Science Center Grant No.
2019/34/E/ST2/00289. M.\'{S}. acknowledge support from the  Project no.
2022/45/P/ST3/04237 co-funded by the National Science Centre, Poland and the
EU’s H2020 research and innovation programme under the MSCA GA no. 945339.

\appendix

\section{Position Measurement} \label{ap:measurement}
We describe our procedure of "measuring"  $N-1$ positions, one by one, from
many-body state $| \Psi \rangle$ to reach the conditional wave-function
$\psi_{\rm cond}$ described in Sec.~\ref{sec:YS}. The procedure is recursive
starting with $| \Psi_1 \rangle = | \Psi \rangle$. Measuring the position
$\bm{r}_1$ of the first particles is done using probability density
\begin{equation}
    p_1(\bm r) \varpropto \langle \Psi_1 | \hat{\psi}^\dag(\bm r) \hat{\psi}(\bm r) | \Psi_1 \rangle.
\end{equation}
Denoting with bar the already drawn position $\bar{\bm{r}}_1$, we then calculate
(not normalized) state with $N-1$ particles as
\begin{equation}
    | \Psi_2 \rangle = \hat{\psi}(\bar{\bm{r}}_1) | \Psi_1 \rangle,
\end{equation}
from which we can draw position of the second particle. Keep in mind that
$\bar{\bm{r}}_1$ is given from previous, step. In general we draw position of
the n-th particle from the distribution
\begin{equation}
    p_n(\bm r) \varpropto \langle \Psi_n | \hat{\psi}^\dag(\bm r) \hat{\psi}(\bm r) | \Psi_n \rangle,
\end{equation}
where
\begin{equation}
    | \Psi_n \rangle = \hat{\psi}(\bar{\bm{r}}_{n-1}) | \Psi_{n-1} \rangle.
\end{equation}
Repeating this procedure, we obtain positions of all $N-1$ particles.

\section{Numerical methods  \label{ap:convergence}}
In this paper we represent the interatomic potential with a narrow Gaussian of
range $\sigma=0.1 R$. We study up to $6$ atoms --  the typical distance between
them is ca. $l=\sqrt{\pi R^2/6}\approx 0.72 R$. As $\sigma$ is substantially
smaller than $l$, the interaction potential is practically a short-range one.
However, we avoid using the Dirac delta potential directly, as it requires proper
regularization in 2D to ensure a finite ground-state energy (see
\cite{Mead1991Oct} and \cite{Mora2003}).

Even with a narrow Gaussian potential, obtaining the ground state necessitates a
big  numerical basis, making brute-force exact diagonalization impractical.
Instead, we employ the importance truncation method (see \cite{Roth2009}) to
construct an optimal many-body basis for diagonalization. In the importance
truncation method one subsequently makes the computational space larger and
larger by repeating the so-called pruning and ranking steps. In these steps, one
simply evaluates the importance of only those Fock states that appear in the
states $\hat{H} |\psi\rangle$, where $|\psi\rangle$ belongs to the computational
Hilbert space obtained in the previous steps. This means that for a given cutoff
$E_{\rm cut}$, our space consists of only a small fraction of all Fock states
with energy smaller than $E_{\rm cut}$. Despite this optimization, our basis
still contains up to 300,000 elements for the strongest interaction strengths
considered. We then diagonalize the Hamiltonian in this basis using the Lanczos
algorithm.

\begin{figure}[!h]
    \centering
    \includegraphics[width=8.6cm]{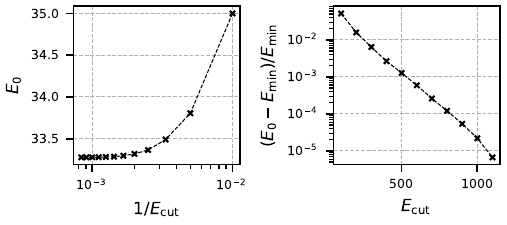}
    \caption{
        Ground state energy as a function of inverse energy cutoff $E_\mathrm{cut}$
        (left) and it's relative error as a function of $E_\mathrm{cut}$ (right).
        The parameter $E_\mathrm{cut}$ is the energy cutoff of the whole Fock space.
        The calculations were performed with importance truncation \cite{Roth2009}
        with parameters $\kappa_\mathrm{min} = 10^{-5}$ and $C_\mathrm{cut} =
        10^{-4}$.
    }
    \label{fig:convergence}
\end{figure}

Figure~\ref{fig:convergence} presents a convergence test for the largest
interaction strength used, $gN = 18\hbar^2/M $. The convergence parameter is the
energy cutoff $ E_\mathrm{cut} $, which determines the Fock states included in
the importance truncation method. The $E_\mathrm{cut}$ is the cutoff for the
maximum total kinetic energy of Fock states. Applying a cutoff on the many-body
Fock states allows us to access these states more effectively than using a
single-particle cutoff.

The left panel shows the ground-state energy
as a function of the inverse energy cutoff, while the right panel displays the
relative error. For the highest energy cutoff used, the relative error remains
below $10^{-5}$, which is sufficient for our qualitative analysis of vortex
emergence and quantum effects.

To conclude, we found that our results are numerically robust.

\end{document}